\documentclass[aps,twocolumn,floats,nofootinbib,prd]{revtex4}
\usepackage{graphicx, epsfig, natbib}

\def\sun{\hbox{$\odot$}}

\begin{document}

\title{Cosmic Acceleration, Dark Energy and Fundamental Physics}
\author{Michael S. Turner and Dragan Huterer}
\affiliation{Kavli Institute for
Cosmological Physics, Department of Astronomy and Astrophysics, and Department
of Physics, The University of Chicago, Chicago, IL 60637}

\begin{abstract}
A web of interlocking observations has established that the expansion of the
Universe is speeding up and not slowing, revealing the presence of some form of
repulsive gravity.  Within the context of general relativity the cause of
cosmic acceleration is a highly elastic ($p\sim -\rho$), very smooth form of
energy called ``dark energy'' accounting for about 75\% of the Universe.  The
``simplest'' explanation for dark energy is the zero-point energy density
associated with the quantum vacuum; however, all estimates for its value are
many orders-of-magnitude too large.  Other ideas for dark energy include a very
light scalar field or a tangled network of topological defects.  An alternate
explanation invokes gravitational physics beyond general relativity.
Observations and experiments underway and more precise cosmological
measurements and laboratory experiments planned for the next decade will test
whether or not dark energy is the quantum energy of the vacuum or something
more exotic, and whether or not general relativity can self consistently
explain cosmic acceleration.  Dark energy is the most conspicuous example of
physics beyond the standard model and perhaps the most profound mystery in all
of science.
\end{abstract}

\maketitle

\section{Quarks and the Cosmos}

The final twenty-five years of the 20th century saw the rise of two highly
successful mathematical models that describe the Universe at its two extremes,
the very big and the very small.  The standard model of particle physics
(detailed in this volume) provides a fundamental description of almost all
phenomena in the microscopic world.  The standard hot big bang model describes
in detail the evolution of the Universe from a fraction of a second after the
beginning, when it was just a hot soup of elementary particles, to the present
some 13.7 billion years later when it is filled with stars, planets, galaxies,
clusters of galaxies and us \cite{Kolb_Turner}.  Both standard models are
consistent with an enormous body of precision data, gathered from high-energy
particle accelerators, telescopes and laboratory experiments.  The standard
model of particle physics and the hot big bang cosmology surely rank among the
most important achievements of 20th century science (see Fig.~1).

Both models raise profound questions.  Moreover, the ``big questions'' about the very small
and the very large are connected, both in their asking and ultimately
in their answering.  This suggests that the deeper understanding that
lies ahead will reveal even more profound connections between the
quarks and the cosmos.  The big questions include

\begin{itemize}
\item How are the forces and particles of nature unified?

\item What is the origin of space, time and the Universe?

\item How are quantum mechanics and general relativity reconciled?

\item How did the baryonic matter arise in the Universe?

\item What is the destiny of the Universe?

\item What is the nature of the dark matter that holds the Universe
together and of the dark energy that is causing the expansion of
the Universe to speed up?
\end{itemize}

The last question illustrates the richness of the connections between quarks
and the cosmos: 96\% of the matter and energy that comprises the Universe
is still of unknown form, is crucial to its existence, and determines
 its destiny.  Dark matter and dark energy are
also the most concrete and possibly most important
evidence for new physics beyond the standard model of particle physics.

The solution to the dark matter problem seems within reach: we have a
compelling hypothesis, namely that it exists in the form of stable elementary
particles left over from the big bang; we know that a small amount of dark
matter exists in the form of massive neutrinos; we have two good candidates for
the rest of it (the axion and neutralino) and an experimental program to test
the particle dark matter hypothesis \cite{Spooner}.

The situation with cosmic acceleration and dark energy is very different.
While we have compelling evidence that the expansion of the
Universe is speeding up, we are far from a working hypothesis or any
significant understanding of cosmic acceleration.  The solution to this
profound mystery could be around the corner or very far away.

\section{Evidence for Cosmic Acceleration}

\subsection{Cosmology Basics}

For mathematical simplicity Einstein assumed that the Universe is isotropic
and homogeneous; today, we have good evidence that this is the case on scales
greater than 100 Mpc (from the distribution of galaxies in the Universe)
and that it was at early times on all scales (from the uniformity of the cosmic
microwave background).  Under this assumption, the expansion is merely a rescaling
and is described by a single function, the cosmic scale factor, $R(t)$.  (By
convention, the value of the scale factor today is set equal to 1.)  The wavelengths of
photons moving through the Universe scale with $R(t)$, and the
redshift that light from a distant object suffers,
$1 + z = \lambda_{\rm rcvd}/\lambda_{\rm emit}$, directly reveals the
size of the Universe when that light was emitted:  $1 + z = 1/R(t_{\rm emit})$.

The key equations of cosmology are
\begin{eqnarray}
H^2 \equiv (\dot R /R)^2 & = & {8\pi G\rho\over 3} -{k\over R^2} +{\Lambda\over 3} \\[0.1cm]
\ddot R/R                & = & -{4\pi G\over 3}\,(\rho + 3p) + {\Lambda\over 3} \\[0.1cm]
w_i \equiv {p_i\over \rho_i}\ \ & \ &\ \ \ \rho_i  \propto   (1+z)^{3(1+w_i)}       \\[0.1cm]
q (z)                   &\equiv & -{\ddot R\over R H^2} = {1\over 2} \left (1+ 3w \right ),
\end{eqnarray}
where $\rho$ is the total energy density of the Universe (sum of matter,
radiation, dark energy) and $p$ is the total pressure. For each component the
ratio of pressure to energy density is the equation-of-state $w_i$ which, through the
conservation of energy, $d(R^3\rho ) = -pdR^3$, determines how the energy density
evolves.  For constant $w$, $\rho \propto (1+z)^{3(1+w)}$:  For
matter ($w=0$) $\rho_M \propto (1+z)^{3}$ and for radiation ($w=1/3$) $\rho_R \propto
(1+z)^4$.  The first of these equations, known as the Friedmann equation,
is the master equation of cosmology.

The quantity $k$ is the 3-curvature of the Universe and $R_{\rm curv}\equiv
R/\sqrt{|k|}$ is the curvature radius; $k = 0$ corresponds to a spatially flat
Universe, $k>0$ a positively curved Universe and $k<0$ a negatively curved
Universe.  Because of the evidence from the cosmic microwave background that
the Universe is spatially flat (see Fig.~1), unless otherwise noted we shall assume $k=0$.

$\Lambda$ is Einstein's infamous cosmological constant; it is equivalent to a
constant energy density, $\rho_\Lambda = \Lambda / 8\pi G$, with pressure
$p_\Lambda = - \rho_\Lambda$ ($w = -1$).  The quantity $q(z)$ is the
deceleration parameter, defined with a minus sign so that $q > 0$ corresponds
to decelerating expansion.

The energy density of a flat Universe ($k=0$), $\rho_C \equiv 3H^2/8\pi G$, is
known as the critical density. For a positively curved Universe, $\Omega_{\rm
TOT} \equiv \rho /\rho_C > 1$ and for a negatively curved Universe $\Omega_{\rm
TOT} <1$. Provided the total pressure is greater than $-1/3$ times the total
energy density, gravity slows the expansion rate, i.e., $\ddot R < 0$ and
$q>0$.  Because of the $(\rho + 3 p)$ term in the $\ddot R$ equation (Newtonian
gravity would only have $\rho$), the gravity of a sufficiently elastic form of
energy ($p < -\rho/3$) is repulsive and causes the expansion of the Universe to
accelerate. In Einstein's static solution ($H = 0$, $q = 0$) the repulsive
gravity of $\Lambda$ is balanced against the attractive gravity of matter, with
$\rho_\Lambda = \rho_M/2$ and $R_{\rm curv}= 1/\sqrt{8\pi G \rho_M}$.  A
cosmological constant that is larger than this results in accelerated expansion
($q < 0$); the observed acceleration requires $\rho_\Lambda \simeq $
(2-3)$\rho_M$.

For an object of known intrinsic luminosity $L$, the measured
energy flux $F$ defines the luminosity distance $d_L$ to the object (i.e.,
the distance inferred from the inverse square law).
The luminosity distance is related to the cosmological model via
\begin{equation}
         d_L(z) \equiv \sqrt{L /4 \pi F} =  (1+z) \int_0^z {dz'\over H(z')}.
\end{equation}
Astronomers determine the luminosity distance from the difference between
the apparent magnitude $m$ of the object (proportional to the
log of the flux) and the absolute magnitude $M$ (proportional to the
log of the intrinsic luminosity), $m-M=5\log_{10}(d_L/10\,{\rm pc})$ (where 5
astronomical magnitudes correspond to a factor of 100 in flux or a
factor of 10 in luminosity distance).

The use of ``standard candles'' (objects of known intrinsic luminosity $L$) and
measurements of the energy flux $F$ constrain the cosmological model through this
equation.  In particular, the Hubble diagram (or magnitude-redshift diagram) is the
simplest route to probing the expansion history.  In
terms of the deceleration parameter the equation is deceptively simple:
\begin{equation}
 H_0 d_L = z + {1\over 2} (1-q_0)z^2 + \cdots
\end{equation}
where the subscript ``0'' denotes the value today.  While this Taylor
expansion of Eq. (5), valid for $z\ll 1$, is of historical significance and
utility, it is not useful today since objects as distant as redshift $z \sim 2$
have been used to probe the expansion history.  However, it does illustrate the
general principle: the first term on the r.h.s. represents the linear Hubble
expansion, and the deviation from a linear relation reveals the deceleration
(or acceleration).

\begin{figure}[t]
\psfig{file=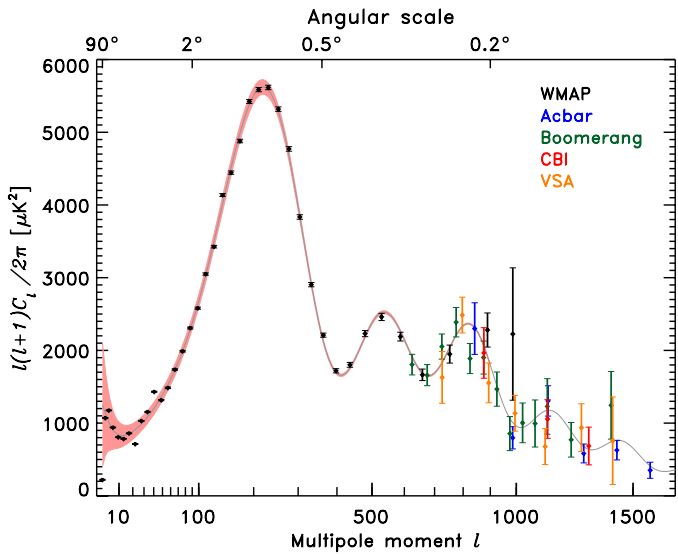,width=3.5in}
\caption{Multipole power spectrum of the CMB temperature fluctuations from
WMAP and other CMB anisotropy experiments.  Position of the first peak at
$l\simeq 200$ indicates the flatness of the Universe; height of the first peak
determines the matter density, and the ratio of the first to second peaks
determines the baryon density.  Together with SDSS large-scale structure data,
the CMB measurements have determined the shape and composition of the Universe:
$\Omega_{\rm TOT} = 1.003\pm 0.010$, $\Omega_M = 0.24 \pm 0.02$, $\Omega_B =0.042\pm
0.002$, and $\Omega_\Lambda = 0.76\pm 0.02$ \cite{LRG}.  The curve is the
theoretical prediction of the ``concordance cosmology'', with a band that
indicates cosmic variance.  Figure adopted from Ref.~\cite{WMAP_website}.  }
\label{fig:CMB}
\end{figure}

\subsection{$\Lambda$'s Checkered History}

Before discussing the evidence for cosmic acceleration, we will recount some of
the history of the cosmological constant.  Realizing that there was nothing to
forbid such a term and that it could be used to obtain an interesting solution
(a static and finite Universe), Einstein introduced the cosmological constant
in 1917.  While his static solution was consistent with astronomical
observations at that time, Hubble's discovery of the expansion of the Universe
in 1929 led Einstein to discard the cosmological constant in favor of expanding
models without one, calling the cosmological constant ``my greatest blunder''.

In 1948, Bondi, Gold and Hoyle put forth the ``steady state cosmology'', with
$\rho_\Lambda > 0$ and $\rho_M \simeq 0$.  The model was motivated by the
aesthetics of an unchanging universe and a serious age problem (the measured
value of the Hubble constant at the time, around 500 km/s/Mpc implied an
expansion age of only 2 Gyr, less than the age of Earth).
The redshift distribution of radio galaxies, the
absence of quasars nearby and the discovery of the cosmic microwave background
radiation in 1960s all indicated that we do not live in an unchanging Universe and
ended this revival of a cosmological constant.

The cosmological constant was briefly resurrected in the late 1960s by
Petrosian et al \cite{Petrosian} to explain the preponderance of quasars at
redshifts around $z \sim 2$ (as it turns out, this is a real effect: quasar
activity peaks around $z \sim 2$).  In 1975 weak evidence for a cosmological
constant from a Hubble diagram of elliptical galaxies extending to redshifts of
$z \sim 0.5$ was presented \cite{Gunn_Tinsley}.  Significant concerns
about whether or not elliptical galaxies were good
standard candles led to the demise of $\Lambda$ once again.  Shortly thereafter came
the rise of the standard cosmology with $\Lambda = 0$.

The current attempt at introducing a cosmological constant (or something
similar), which is backed up by multiple lines of independent evidence, traces
its roots to the inflationary universe scenario and its prediction of a
spatially flat Universe.  In the early 1980s when inflation was introduced, the
best estimate of the average mass density fell short of the critical density by
almost a factor of ten ($\Omega_M \sim 0.1$); the saving grace for inflation
was the large uncertainty associated with measuring the mean matter density.
From 1980 to the mid 1990s, as measurement techniques took better account of
dark matter, $\Omega_M$ rose to of order 0.5 or so.  However, as the
uncertainties got smaller, $\Omega_M$ began converging on a value of around
1/3, not 1.  Moreover, the predictions of the cold dark matter scenario of
structure formation matched observations if $\Omega_M$ was around 1/3, not 1.

Starting in 1984 and continuing to just before the discovery of cosmic
acceleration, a number of papers suggested the solution to inflation's
``$\Omega$ problem'' \cite{Kolb_Turner} was a cosmological constant
\cite{Omega_prob}.  Owing to its checkered history, there was not much
enthusiasm for this suggestion at first.  However, with time the indirect
evidence for $\Lambda$ grew \cite{Krauss_Turner,JPO_PJS,Turner_White}, and in
1998 when the supernova evidence for accelerated expansion was presented the
cosmological constant was quickly embraced -- this time, it was the piece of
the puzzle that made everything work.

\subsection{Discovery and confirmation}

\begin{figure}[t]
\psfig{file=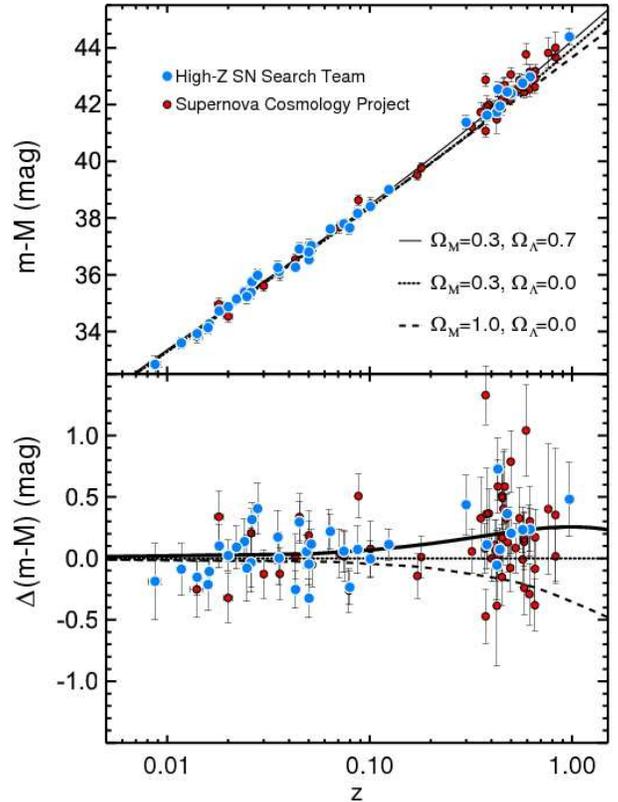,width=3.7in}
\caption{Discovery data: Hubble diagram of SNe
Ia measured by the Supernova Cosmology Project and High-z Supernova Team. Bottom
panel shows magnitudes relative to a universe with
$\Omega_{\rm TOT} = \Omega_M=0.3$. Figure adopted from Ref.~\cite{Perlmutter_Schmidt}.  }
\label{fig:hub_diag}
\end{figure}

Two breakthroughs enabled the discovery that the Universe is
speeding up and not slowing down.  The first was the demonstration that type Ia
supernovae (SNe Ia), the brightest of the supernovae and the ones believed to
be associated with the thermonuclear explosions of 1.4 $M_{\sun}$ white-dwarf
stars pushed over the Chandrasekhar mass limit by accretion, are (nearly) standard candles
\cite{Phillips}.  The second breakthrough involved the use of large (of order
100 Megapixel) CCD cameras to search big regions of the sky containing
thousands of galaxies for these rare events (the SN Ia rate in a typical galaxy
is of the order of one per 100 to 200 years).  By comparing images of thousands
of galaxies taken weeks apart the discovery of SNe could be reliably
``scheduled'' on a statistical basis.

\begin{figure}[t]
\psfig{file=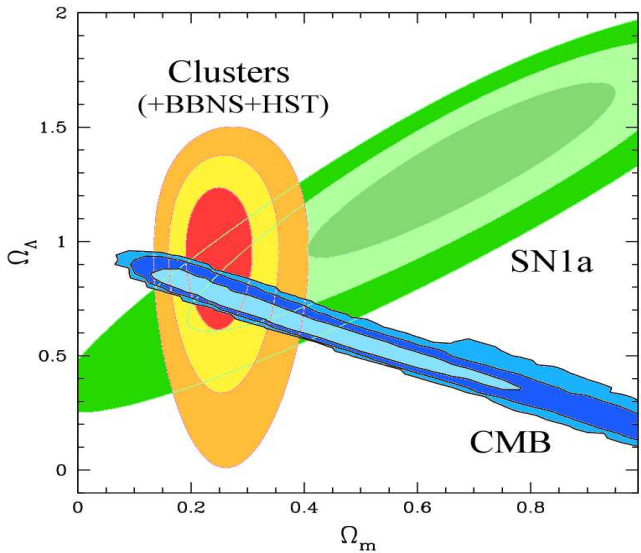,width=3.5in, height=3.0in}
\caption{Three independent lines of evidence for cosmic acceleration:
galaxy clusters, CMB anisotropy and SNe.  Flatness is not assumed,
but $w=-1$. Note the concordance of the independent methods; evidence for
$\Omega_\Lambda > 0$ is greater 99.9\% C.L. Figure adopted from Ref.~\cite{allen}.
}
\label{fig:om_ol}
\end{figure}

Two teams working independently in the mid- to late-1990s used
took advantage of these breakthroughs to determine the expansion
history of the Universe.  They both found that distant
SNe are dimmer than they would be in a decelerating Universe, indicating that
the expansion has actually been speeding up for the past 5 Gyr
\cite{Riess_98, Perlmutter_99}; see Fig.~\ref{fig:hub_diag}. Analyzed for a Universe
with matter and cosmological constant, their results provide evidence for
$\Omega_{\Lambda}>0$ at greater than 99\% confidence; see Fig.~\ref{fig:om_ol}.

Since this work, the two teams have discovered and studied more SNe, as
have other groups \cite{Knop,Riess_04_06,SNLS,ESSENCE}.  Not only has the
new data confirmed the discovery, but it has also allowed measurements
of the equation-of-state of dark
energy $w=p/\rho$ (assuming constant $w$), and even constrains the time
variation of $w$, with the parametrization $w = w_0 + w_a(1-R)$.

\begin{figure}[!t]
\psfig{file=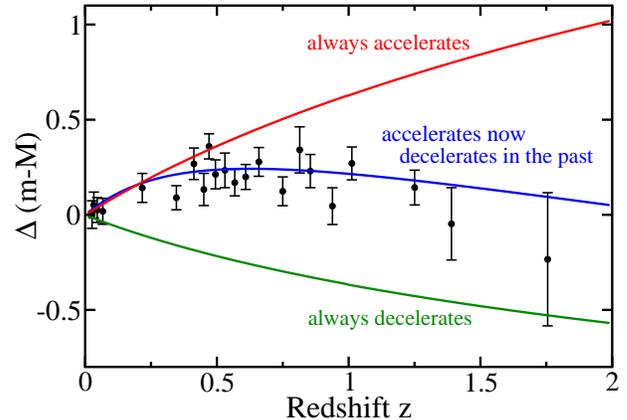,height=3.5in, angle=-90}
\caption{Evidence for transition from acceleration today
to deceleration in the past. The Hubble diagram with measured
distances to SNe Ia is not fitted well by either a purely accelerating or
purely decelerating cosmological model, but rather one with early deceleration
and recent acceleration. SN data are binned in redshift and come from
Ref.~\cite{Riess_04_06}.  }
\label{fig:deceleration}
\end{figure}

Especially important in this regard are SNe with redshifts $z>1$
which indicate that the universe was decelerating at earlier times (see
Fig.~\ref{fig:deceleration}), and hence that dark energy started its domination
over the dark matter only recently, at redshift $z=(\Omega_M/\Omega_{\rm
DE})^{1/3w}-1\approx 0.5$.  This finding is an important reality check:
without a long, matter-dominated, slowing phase, the Universe could not have
formed the structure we see today.

Evidence for dark energy comes from several other independent
probes. Measurements of the fraction of X-ray emitting gas to total mass in
galaxy clusters, $f_{\rm gas}$, also indicates the presence of dark energy.
Because galaxy clusters are the largest collapsed objects in the universe, the
gas fraction in them is presumed to be constant and equal to the overall baryon
fraction in the universe, $\Omega_B/\Omega_M$ (most of the baryons in clusters
reside in the gas).  Measurements of the gas fraction $f_{\rm gas}$ depend not
only on the observed X-ray flux, but also on the distance to the cluster;
therefore, only the correct cosmology will produce distances which make the
apparent $f_{\rm gas}$ constant in redshift. Using data from the Chandra X-ray
Observatory, Allen et al \cite{allen} have determined $\Omega_\Lambda$ to an
accuracy of about $\pm 0.2$; see Fig.~\ref{fig:om_ol}.

Cosmic microwave background (CMB) anisotropies provide a record of the Universe
at simpler time, before structure had developed and when photons were decoupling
from baryons $z\approx 1100$ \cite{Hu_Dodelson}. The multipole power
spectrum is dominated by the acoustic peaks that arise from gravitationally
driven photon-baryon oscillations (see Fig.~\ref{fig:CMB}).  The positions and
amplitudes of the acoustic peaks encode much information about the Universe,
today and at earlier times.  In particular, they indicate that the Universe is
spatially flat, with a matter density that accounts
for only about a quarter of the critical density.  However, the presence
of a uniformly distributed energy density with large negative
pressure which accounts for three-quarters of the critical density
brings everything into good agreement, both with CMB data and the large-scale
distribution of galaxies in the Universe.
The CMB data of WMAP together with large-scale structure data of
the Sloan Digital Sky Survey (SDSS) provides the following cosmic census
\cite{LRG}: $\Omega_{\rm TOT} = 1.003\pm 0.010$, $\Omega_M = 0.24 \pm 0.02$,
$\Omega_B =0.042\pm 0.002$, and $\Omega_\Lambda = 0.76\pm 0.02$.

The presence of dark energy affects the large-angle
anisotropy of the CMB (the low multipoles) and leads to the prediction of
a small correlation between the galaxy distribution and the CMB anisotropy.
This subtle effect has been observed \cite{ISW_cross}; while not detected
at a level of significance that could be called independent confirmation,
its presence is a reassuring cross check.

\begin{figure}[t]
\psfig{file=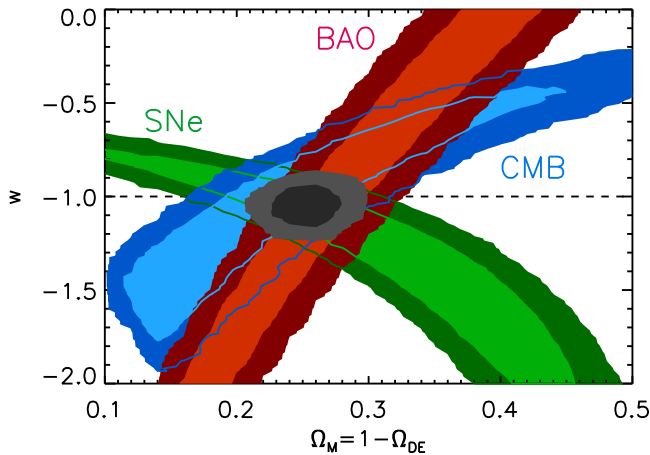,width=3.5in}
\caption{68\% and 95\% C.L. constraints on the matter density $\Omega_M$ and dark
energy equation-of-state $w$, assuming a flat Universe.  Constraints come from
the Supernova Legacy Survey \cite{SNLS}, WMAP \cite{WMAP}, and SDSS detection of
BAO \cite{BAO}.  The combined constraint is shown by the central dark
contours.}
\label{fig:om_w}
\end{figure}

Baryon acoustic oscillations (BAO), so prominent in the CMB anisotropy (see Fig.~1),
leave a smaller characteristic signature in the clustering of galaxies that can be
measured today and provide an independent geometric probe of dark energy.
Measurements of the BAO signature in the correlation function
of SDSS galaxies constrains the distance to redshift $z=0.35$
to a precision of 5\% \cite{BAO}.  While this alone does not establish the
existence of dark energy, it serves as a significant complement to other probes,
cf.,\ Fig.~\ref{fig:om_w}.

Weak gravitational lensing \cite{WL_reviews} -- slight distortions of galaxy
shapes due to gravitational lensing by intervening large-scale structure -- is
a powerful technique for mapping dark matter and its clustering. Currently,
weak lensing sheds light on dark energy by pinning down the combination
$\sigma_8 (\Omega_M/0.25)^{0.6}\approx 0.85\pm 0.07$, where $\sigma_8$ is the
amplitude of mass fluctuations on the 8\,Mpc scale \cite{WL}.  Since other
measurements put $\sigma_8$ at $\sim 0.9$, this implies that $\Omega_M \simeq
0.25$, consistent with a flat Universe whose mass/energy density is dominated
by dark energy. In the future, weak lensing will also be very useful in probing
the equation-of-state of dark energy \cite{Huterer_WL}; see Sec.~IV.

\begin{figure}[t]
\psfig{file=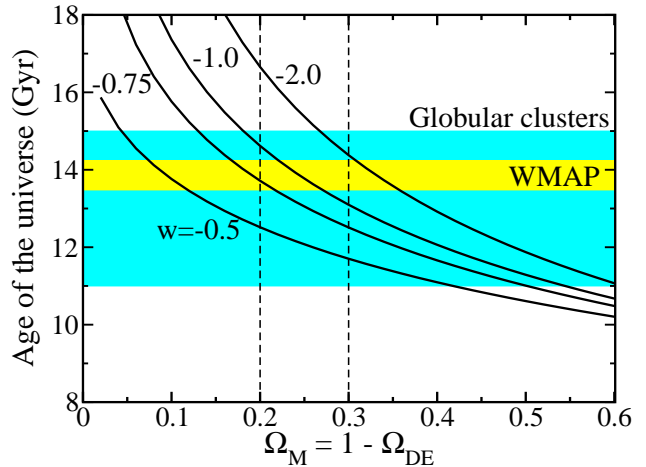, height=3.5in, angle=-90}
\caption{Age of the universe as a function of the matter energy density,
assuming a flat universe and four different values of the dark
energy equation-of-state. Also shown are constraints from globular
clusters \cite{Krauss_Chaboyer} and from WMAP \cite{WMAP} and the
range of $\Omega_M$ favored by measurements of the matter density. Age consistency
holds for $-2\lesssim w\lesssim -0.75$.}
\label{fig:age}
\end{figure}

Finally, because the expansion age of the Universe, $t_0 = \int\,dz/(1+z)H(z)$,
depends upon the expansion history, the comparison of this age with other
independent age estimates can be used to probe dark energy.  The ages of the
oldest stars in globular clusters constrain the age of the Universe: $11\,{\rm
Gyr}\lesssim t_0\lesssim 15\,{\rm Gyr}$
\cite{Krauss_Chaboyer}. CMB anisotropy is very sensitive to
the age, and WMAP data determine it accurately:
$t_0=13.84^{+0.39}_{-0.36}\,{\rm Gyr}$ \cite{WMAP}.  Fig.~\ref{fig:age}
shows that a consistent age is possible if $-2\lesssim w\lesssim -0.75$.
Agreement on the age of the Universe provides an important consistency check 
as well as confirmation of a key feature of dark energy, 
its large negative pressure.

\section{Understanding Cosmic Acceleration}

Sir Arthur Eddington is quoted as saying, ``It is (also) a good rule not to put
too much confidence in observational results until they are confirmed by
theory''. While this may seem a bit paradoxical (or worse yet, an example of
blatant theoretical arrogance), the point is well taken: science is not just a
collection of facts, it is also understanding; if the understanding does not
eventually follow new facts, perhaps there is something wrong with the facts.

Cosmic acceleration meets the Eddington criterion and at the same time 
presents a stunning opportunity for theorists:
General relativity (GR) can accommodate accelerated
expansion, but GR has yet to provide a deeper understanding of
the phenomenon.

Within GR, a very elastic fluid has repulsive gravity,
and, if present in sufficient quantity, can lead to the observed accelerated expansion.
This then is the definition of dark energy: the mysterious, elastic and
very smooth form of energy which is responsible for cosmic acceleration and
is characterized by an equation-of-state $w=p/\rho \sim -1$ \cite{Turner_White}.

Vacuum energy is a concrete example of dark energy.  General
covariance requires that the stress energy
associated with the vacuum take the form of a constant times
the metric tensor.  This implies that it has a pressure equal to minus
its energy density, is constant both in space and time,
and is mathematically equivalent to a cosmological constant.

The stress energy associated with a homogeneous scalar
field $\phi$ can also behave like dark energy.  It
takes the form of a perfect fluid with
\begin{equation}
\rho = \dot{\phi}^2/2  + V(\phi ) \ \ \ \ \ \ \ p    = \dot{\phi}^2/2  - V(\phi ),
\end{equation}
where $V(\phi )$ is the potential energy of the scalar field, dot denotes time
derivative, and the evolution of the field $\phi$ is governed by
\begin{equation}
  \ddot {\phi} + 3 H\phi + V'(\phi) = 0.
\end{equation}
If the scalar field evolves slowly, that is $\dot{\phi}^2 \ll V$, then $p
\simeq -\rho$ and the scalar field behaves like a slowly varying vacuum energy.

While cosmic acceleration can be accommodated within the GR framework, the
fundamental explanation could be new gravitational physics.  With this as
a prelude, we now briefly review the present theoretical situation.

(a) {\it Vacuum energy.}\quad  Vacuum energy is both the most plausible
explanation and the most puzzling possibility.  For almost 80 years we have
known that there should in principle be an energy associated with the
zero-point fluctuations of all quantum fields.  Moreover, $p_{\rm VAC} = -\rho_{\rm
VAC}$.  However, all attempts to compute the value of the vacuum
energy lead to divergent results.  The so-called cosmological constant problem was
finally articulated about thirty years ago \cite{Weinberg}.  However,
because of the success of the standard hot big bang model (where $\Lambda = 0$)
and the absence of good (or any) ideas, the problem was largely ignored.
With the discovery of cosmic acceleration, the cosmological constant problem
is now front and center and can no longer be ignored.

To be more quantitative, the energy density required to explain the accelerated
expansion is about three quarters of the critical density or about $4\times 10^{-47}
{\rm GeV}^4 \approx (3\times 10^{-3} {\rm eV})^4$.  This is tiny compared to
energy scales in particle physics (with the exception of
neutrino mass differences).  Such a small energy precludes solving the problem
by simply cutting off the divergent zero-point energy integral at some energy
beyond which physics is not yet known.  For example, a cutoff of
100 GeV would leave a 54 orders-of-magnitude discrepancy.  If supersymmetry
were an unbroken symmetry, fermionic and bosonic zero-point
contributions would cancel.   However, if supersymmetry is broken at a scale of
order $M$, one would expect that imperfect cancellations leave a
finite vacuum energy of the order $M^4$, which for the favored value of $M \sim
100$ GeV to $1$ TeV, would leave a discrepancy of 50 or 60 orders-of-magnitude.

One approach to the cosmological constant problem involves the idea
that the value of the vacuum energy is a random variable which can take on
different values in different disconnected pieces of the Universe.  Because a
value much larger than needed to explain the observed cosmic acceleration would
preclude the formation of galaxies, we could not find ourselves in such a
region \cite{Efstathiou}.  This very anthropic
approach finds a home in the landscape version of
string theory \cite{Susskind}.

(b) {\it Scalar fields, etc.}\quad While introducing a new dynamical degree of
freedom can also provide a very elastic form of energy density, it does not
solve the cosmological constant problem.  In order to roll slowly enough the
mass of the scalar field must be very light, $m \lesssim H_0 \sim
10^{-42}$\,GeV, and its coupling to matter must be very weak to be consistent
with searches for new long-range forces \cite{Carroll_quint}. Unlike vacuum
energy, scalar-field energy clusters gravitationally, but only on the largest
scales and with a very small amplitude \cite{Hu_GDM}.

The equation-of-state $w$ for a scalar field can take on any value between $-1$
and $1$ and in general varies with time.  (It is also possible to have $w <
-1$, though at the expense of ghosts, by changing the sign of the kinetic
energy term in the Lagrangian.)  Scalar field models also raise new questions
and possibilities: Is cosmic acceleration related to inflation?  Is dark energy
related to dark matter or neutrino mass?  No firm or compelling connections
have been made to either, although interesting possibilities have been
suggested.

Scalar fields in a very different form can also explain cosmic acceleration.
The topological solitons that arise in broken gauge theories, e.g., strings,
walls and textures, are very elastic, and tangled networks of such defects can
on large scales behave like an elastic medium with $w = -N/3$, where $N$ is the
dimensionality of the network ($N=1$ for strings, 2 for walls and 3 for
textures).  In this case $w$ is a fixed, rational number.

(c) {\it Modified gravity.}\quad A very different approach holds that cosmic
acceleration is a manifestation of new gravitational physics and not of dark
energy.  Assuming that our 4-d spacetime can still be described by a metric,
the operational changes are twofold: (1) a new version of the Friedmann
equation governing the evolution of the background spacetime; (2) modifications
to the equations that govern the growth of the small matter perturbations that
evolve into the structure seen in the Universe today.  A number of ideas have
been explored, from models motivated by higher-dimensional theories and string
theory \cite{DGP,high_D} to generic modifications of the usual gravitational
action \cite{f(R)}.

An aside:  One might be concerned that when the assumption of general relativity is
dropped the evidence for accelerated expansion might disappear.  This is
not the case; using the deceleration $q(z)$ as a kinematic description of the
expansion, the SNe data still provide strong
evidence for a period of accelerated expansion \cite{Shapiro_Turner}.

Changes to the Friedmann equation are easier to derive, discuss and
analyze.  In order not to spoil the success of the standard cosmology at
early times (from big-bang nucleosynthesis to the CMB anisotropy to the
formation of structure), the Friedmann equation must reduce to the GR
form for $z \gg 1$.  Because the matter term scales as $(1+z)^3$
and the radiation term as $(1+z)^4$, to be safe any modifications should
decrease with redshift more slowly than this.  As a specific example, 
consider the DGP model, which arises from a five-dimensional gravity 
theory \cite{DGP}, and has a 4-d Friedmann equation,
\begin{equation}
  H^2 = {8\pi G \rho_M \over 3} + {H\over r_c},
\end{equation}
where $r_c$ is an undetermined scale.  As $\rho_M \rightarrow 0$, there is a
(self) accelerating solution, with $H= 1/r_c$.  The additional term in the
Friedmann equation, $H/r_c$, behaves just like dark energy with an equation-of-state
that evolves from $w=-1/2$ (for $z \gg 1$) to $w = -1$ in the distant future.

\section{Prospects for revealing the nature of Dark Energy}

We divide the probes of dark energy into three broad categories: kinematical
and dynamical cosmological probes, and laboratory/astrophysical probes.
Kinematical tests rely on the measurement of cosmological distances and volumes
to constrain the evolution of the scale factor and thus the background
cosmological model. Specific techniques include SNe Ia, CMB, and baryon
acoustic oscillations.

The dynamical tests probe the effect of dark energy on perturbations of the
cosmological model, including the evolution of the small inhomogeneities in the
matter density that give rise to structure in the Universe.  Specific
techniques include the use of gravitational lensing to directly determine the
evolution of structure in the dark matter and the study of the growth of the
abundance of galaxy clusters to indirectly probe the growth of structure.  A
potential probe of dark energy, which at the present seems beyond reach, is to
study the clustering of dark energy itself.  Since vacuum energy does not
cluster, detection of such would rule out vacuum energy as the explanation for
cosmic acceleration.

In general relativity, for both kinematical and dynamical
cosmological probes, the primary effect of dark energy enters through the
Friedmann equation, cf. Eq. (1),
\begin{eqnarray}
H(z)^2  & = & {8\pi G \over 3} \left[ \rho_M + \rho_{DE} \right] \nonumber \\[0.1cm]
        & = & H_0^2 \left[ \Omega_M (1+z)^3 + (1-\Omega_M)(1+z)^{3(1+w)}
\right],
\end{eqnarray}
where a flat Universe and constant $w$ have been assumed.  In turn, the
expansion rate affects the luminosity distance, $d_L = (1+z)\int\, dz/H(z)$,
the number of objects seen on the sky, $d^2 N/(d\Omega\, dz) = n(z)\,
d_L^2 /[(1+z)^2H(z)]$ ($n$ is the comoving density of objects), and the
evolution of cosmic structure via the growth of small density perturbations.
In GR the growth of small density perturbations in the matter is governed by
\begin{equation}
\ddot\delta_k + 2H\dot\delta_k -4\pi G \rho_M \delta_k = 0\, ,
 \label{eq:growth}
\end{equation}
where density perturbations in the cold dark matter have been decomposed into
their Fourier modes of wavenumber $k$.  Dark energy affects the growth through
``the drag term'', $2H\dot\delta_k$.  The equations governing dark energy
perturbations depend upon the specific dark energy model.

The kinematical and dynamical tests probe complementary aspects of the effect of dark
energy on the Universe:  the overall expansion of the Universe (kinematical) and
the evolution of perturbations (dynamical).  Together,
they can test the consistency of the underlying gravity
theory.  In particular, different values of
the dark energy equation-of-state obtained by the two methods would
indicate an inconsistency of the underlying gravity theory.

The kinematical tests are easier to frame because they only depend upon knowing
the effect of dark energy on the background cosmological model.  Further,
cosmological variables (such as $q$) can even be formulated without
reference to a particular theory of gravity.  The dynamical tests are both harder
to frame --- they require detailed knowledge of how dark energy clusters and
affects the growth of density perturbations --- and also harder to implement ---
they rely upon the details of describing and measuring the distribution of
matter in the Universe.  Both the kinematical and dynamical tests have their
greatest probative power at redshifts between about $z = 0.2$ and $z = 2$, for
the simple reason that at higher redshifts dark energy becomes increasingly
less important, $\rho_{DE}/ \rho_M \propto (1+z)^{3w}$ \cite{Hut_Tur}.

While the primary probes of dark energy are cosmological, laboratory
experiments may be able to get at the underlying physics.  If dark energy
couples to matter there will be long-range forces that are in principle
detectable; if it couples to electromagnetism, polarized light from distant
astrophysical sources should suffer rotation \cite{Carroll_quint}.  It is also
possible that accelerator-based experiments will have something to say about
dark energy. For example, if evidence for supersymmetry is found at the Large
Hadron Collider, understanding how supersymmetry is broken could shed light
on the vacuum energy puzzle.

\begin{figure}[t]
\epsfig{file=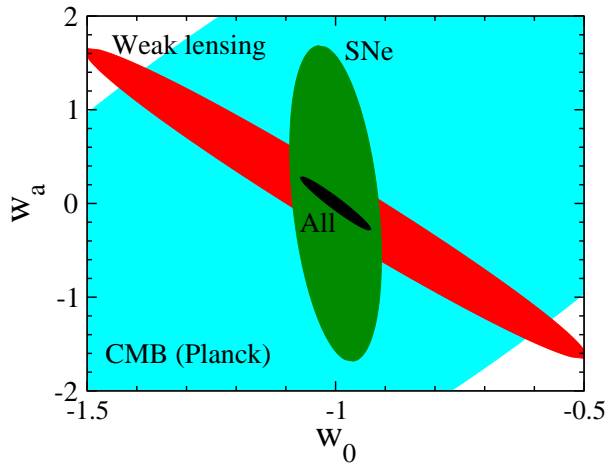, height=3.3in, angle=-90}
\caption{Forecasts of future constraints to the dark energy
equation-of-state and its time evolution using
SNe Ia, weak gravitational lensing, and CMB anisotropy (measured by
Planck).  Future SNe Ia and weak lensing estimates are both modeled on the SNAP
experiment.  The large improvement in combining the various data sets is due to
breaking of the parameter degeneracies in the full (8-dimensional) parameter
space.  }
\label{fig:w0wa}
\end{figure}

Observations to date have established the existence of dark energy and have
begun to probe its nature; e.g., by constraining 
$w\approx -1\pm 0.1$.  Future experiments will
focus on testing whether or not it is vacuum energy and the consistency of GR
to accommodate dark energy. The Supernova/Acceleration Probe (SNAP)
\cite{SNAP}, a proposed space-based telescope to collect several thousand SNe
out to $z\approx 2$, would significantly reduce uncertainties (both statistical
and systematic) on dark-energy parameters.  SNAP, together with the planned
wide-field surveys from the ground, the Dark Energy Survey (DES) \cite{DES} and
the Large Synoptic Survey Telescope (LSST) \cite{LSST}, would map the weak
lensing signal from one arcminute out to the largest observable scales on the
sky and accurately determine the effect of dark energy on the growth of
structure. Large BAO surveys are also planned, both from the ground and
space. The just-completed South Pole Telescope (SPT) \cite{SPT} and the Atacama
Cosmology Telescope (ACT) \cite{ACT} will soon begin studying dark energy by
determining the evolution of the abundance of galaxy clusters.  In 2008, ESA's
Planck Surveyor CMB satellite \cite{Planck} will be launched and will extend
precision measurements of CMB anisotropy to $l\sim 3000$ (i.e.\ down to angular scales
of about 5 arcmin), more accurately
pinning down the matter density and providing
an important prior constraint for other dark energy
measurements \cite{Friemanetal}.
Theoretical forecasts for future constraints on the parameters $(w_0, w_a)$ are
shown in Fig.~\ref{fig:w0wa}.

\section{Dark Energy and Destiny}

\begin{figure}[t]
\psfig{file=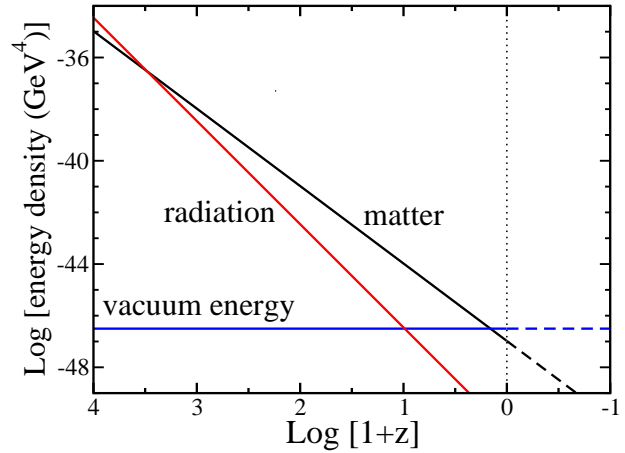,height=3.5in, angle=-90}
\caption{Evolution of dark matter, dark energy and radiation.  Earlier than
$z\sim 3000$ radiation dominates the mass/energy density of the
Universe; between $z\sim 3000$ and $z\sim 0.5$ dark matter dominates,
and thereafter dark energy dominates and the expansion accelerates.  Structure
only grows during the matter dominated epoch.  }
\label{fig:scalings}
\end{figure}

One of the first things one learns in cosmology is that geometry is destiny: a
closed (positively curved) Universe eventually recollapses and an open (flat or negatively
curved) Universe expands forever.  Provided that the Universe only contains
matter and $\Lambda= 0$, this follows directly from Eq.~(1).  If $k>0$, the
Universe achieves a maximum size when $H^2$ is driven to zero by the inevitable
cancellation of $\rho_M$ and $k/R^2$.  If $k=0$, $R$ always grows as $t^{2/3}$
and $q = 1/2$.  For $k<0$, the Universe ultimately reaches a
coasting phase where $R$ grows as $t$ and $q = 0$. Adding radiation only
changes the story at early times (see Fig.~\ref{fig:scalings}): because the
radiation density increases as $(1+z)^4$, for $z\gg 3\times 10^3$ the Universe is
radiation dominated and during this epoch, $R \propto t^{1/2}$ and $q = 1$
(this is a manifestation of the fact that gravity is sourced by $\rho + 3p =
2\rho$ for radiation, cf. Eqns.\ 2 and 4).  It is only during the
matter-dominated phase that small density inhomogeneities are able to grow and
form bound structures.

Dark energy provides a new twist: because the dark energy density varies slowly
if at all, it eventually becomes the dominant form of matter/energy (around $z
\sim 0.5$); see Fig.~8.  After that, the expansion accelerates and structure formation
ceases, leaving in place all the structure that has formed.  The future beyond
the present epoch of accelerated expansion is uncertain and depends upon
understanding dark energy.

In particular, if dark energy is vacuum energy, acceleration will continue and
the expansion will become exponential, leading inevitably to a dark Universe.
(In a hundred billion years, the light from all but a few hundred nearby
galaxies will be too redshifted to detect.)  On the other hand, if dark energy
is explained by a scalar field, then eventually the field relaxes to the
minimum of its potential.  If the minimum of the potential energy is zero, the
Universe again becomes matter dominated and returns to decelerated expansion.
If the minimum of the scalar potential has negative energy density,
the energy of dark matter and of scalar field energy will eventually cancel,
leading to a recollapse.  Finally, if the potential energy at the minimum is
positive, no matter how small, accelerated expansion eventually ensues again.

Absent dark energy geometry and destiny are linked.  The presence of dark energy
severs this relation \cite{Krauss_Turner} and links instead destiny to an
understanding of dark energy.

\section{Summary}

We end our brief review with our list of the ten most important facts
about cosmic acceleration

1. Independent of general relativity and based solely upon the SN Hubble
   diagram, there is very strong evidence that the expansion of the Universe
   has accelerated recently \cite{Shapiro_Turner}.

2. Within the context of general relativity, cosmic acceleration cannot be
   explained by any known form of matter or energy, but can be accommodated by
   a nearly smooth and very elastic ($p \sim -\rho$) form of energy (``dark energy'') that
   accounts for about 75\% of the mass/energy content of the Universe.

3. Taken together, current data (SNe, galaxy clustering, CMB and galaxy clusters) provide
   strong evidence for the existence of dark energy and constrain the
   fraction of critical density contributed by dark energy to be $71\pm 5$\% and
   the equation-of-state to be $w\approx -1 \pm 0.1$ (stat) $\pm 0.1$ (sys), with
   no evidence for variation in $w$.  This implies that the Universe
   decelerated until $z \sim 0.5$ and age $\sim 10$ Gyr, when it began accelerating.

4. The simplest explanation for dark energy is the zero-point energy of the
   quantum vacuum, mathematically equivalent to a cosmological constant.
   In this case, $w$ is precisely $-1$, exactly uniformly
   distributed and constant in time.  All extant data are consistent
   with a cosmological constant; however, all attempts to compute the energy of
   the quantum vacuum yield a result that is
   many orders-of-magnitude too large (or is infinite).

5. There is no compelling model for dark energy.  However there are many
   intriguing ideas including a new light scalar field, a tangled network of
   topological defects, or the influence of additional spatial dimensions.
   It has also been suggested that dark energy is related to cosmic
   inflation, dark matter and neutrino mass.

6. Cosmic acceleration could be a manifestation of gravitational physics beyond
   general relativity rather than dark energy.  While there are intriguing
   ideas about corrections to the usual gravitational action or 
   modifications to the Friedmann equation that can
   give rise to the observed accelerated expansion, there is
   no compelling, self-consistent model for the new gravitational physics that
   explains cosmic acceleration.

7. Even assuming the Universe has precisely the critical density and is
   spatially flat, the destiny of the Universe depends crucially upon the nature of the
   dark energy.  All three fates --- recollapse or continued expansion with and
   without slowing --- are possible.

8. Cosmic acceleration is arguably the most profound puzzle in physics.  Its
   solution could shed light on or be central to
   unraveling other important puzzles, including the cause of cosmic
   inflation, the vacuum-energy problem, supersymmetry and
   superstrings, neutrino mass, new gravitational physics, and dark matter.

9. Today, the two most pressing questions about dark energy and cosmic acceleration are:
   Is dark energy something other than vacuum energy? Does general relativity
   self consistently describe cosmic acceleration?  Establishing that $w \not= -1$
   or that it varies with time would rule out vacuum energy; establishing that
   the values of $w$ determined by the kinematical and dynamical methods are not equal
   would indicate that GR cannot self consistently accommodate accelerated expansion.

10. Dark energy affects the expansion rate of the Universe, which in turn affects
    the growth of structure and the distances to objects.  (In gravity theories other
    than GR, dark energy may have more direct effects on the growth of structure.)
    Upcoming ground- and space-based experiments should probe $w$ at the percent level and its
    variation at the ten percent level.  These measurements should dramatically improve our
    ability to discriminate between vacuum energy and something more exotic as
    well as testing the self consistency of general relativity.  Laboratory- and
    accelerator-based experiments could also shed light on dark energy.

Because of its brevity, this review could not do justice to the extensive
literature that now exists; for readers interested in a more thorough treatment
of dark energy and/or a more extensive review, we refer them to
Refs.~\cite{reviews}.

\end{document}